\documentclass{article}
\usepackage{amsmath,graphicx,mlspconf,hyperref}
\usepackage{xcolor}
\usepackage{amssymb}
\usepackage{textcomp} 
\usepackage{gensymb}
\newcommand\inner[2]{\langle #1, #2 \rangle}
\newcommand{\norm}[1]{\left\lVert#1\right\rVert}
\usepackage{gensymb}

\copyrightnotice{979-8-3503-2411-2/25/\$31.00 {\copyright}2025 European Union}

\toappear{Presented at IEEE International Workshop on Machine Learning for Signal Processing, Aug.\ 31-- Sep.\ 3, 2025, Istanbul, Turkey}

\title{PROTOTYPICAL CONTRASTIVE LEARNING FOR IMPROVED FEW SHOT AUDIO CLASSIFICATION}

\name{Anonymous\thanks{Anonymous.}}
\address{Anonymous}

\name{%
    C. Sgouropoulos,$^{\dagger}$
    C. Nikou,$^{\dagger}$
    S. Vlachos,$^{\dagger}$
    V. Theiou,$^{\dagger}$ 
    C. Foukanelis,$^{\dagger}$
    T. Giannakopoulos$^{\dagger}$%
\thanks{This work was supported by the European Union through the \href{https://faradai.eu}{FaRADAI} Project. Funded by the European Union. Views and opinions expressed are however those of the author(s) only and do not necessarily reflect those of the European Union nor the European Commission. Neither the European Union nor the granting authority can be held responsible for them.}
}

\address{
    $^{\dagger}$ Multimedia Analysis Group of the Computational Intelligence Laboratory (MagCIL)\\
    Institute of Informatics and Telecommunications, NCSR "DEMOKRITOS"
}

\begin{document}
\maketitle
\begin{abstract}
Few-shot learning has emerged as a powerful paradigm for training models with limited labeled data, addressing challenges in scenarios where large-scale annotation is impractical. While extensive research has been conducted in the image domain, few-shot learning in audio classification remains relatively underexplored. In this work, we investigate the effect of integrating supervised contrastive loss into prototypical few shot training for audio classification. In detail, we demonstrate that angular loss further improves the performance compared to the standard contrastive loss. Our method leverages SpecAugment followed by a self-attention mechanism to encapsulate diverse information of augmented input versions into one unified embedding. We evaluate our approach on MetaAudio, a benchmark including five datasets with predefined splits, standardized preprocessing, and a comprehensive set of few-shot learning models for comparison. The proposed approach achieves state-of-the-art performance in a 5-way, 5-shot setting. 
\end{abstract}

\newcommand{\cem}[1]{\textcolor{blue}{cem: #1}}
\section{Introduction}
\label{sec:intro}

In today's rapidly evolving fields of machine learning and artificial intelligence, there is a growing demand for models that can generalize effectively from limited training data. Traditional machine learning algorithms typically depend on large amount of labeled data to achieve high performance, making them less effective in real-world scenarios where such data are often scarce or difficult to obtain. In contrast, Few-Shot Learning (FSL) focuses on enabling models to achieve high performance with only a few labeled examples. This approach is especially valuable in scenarios where the data is limited or unevenly distributed, allowing models to quickly adapt to new tasks with minimal prior information.

Existing methodologies in few-shot learning can be broadly categorized into metric learning and optimization-based methods. Metric learning approaches utilize well-defined similarity measures to compare query samples with support examples, enabling classification based on embedding proximity. These methods often use techniques such as ProtoNets and MatchingNets~\cite{Protonets,Matchingnets}, to learn discriminative embeddings by utilizing various distance metrics. On the other hand, optimization-based methods focus on efficiently adapting model parameters based on limited data, leveraging meta-learning strategies such as gradient-based updates~\cite{MAML,MAMLcurva} or task-specific adaptation mechanisms \cite{TADAM}.

Although extensive research has been conducted in the image domain, few-shot learning in the audio domain remains relatively underexplored. Several studies focus on the task of sound event detection, aiming to identify specific acoustic events in an audio file using only a few examples~\cite{fewshotsoundevent,Transformerbio}.
In the context of few shot audio classification, researchers have leveraged Prototypical Networks for the tasks of speaker recognition~\cite{centroid} and sound event classification~\cite{audio_classif}. Chou et al.~\cite{transient} incorporate an attention-based similarity mechanism into metric learning architectures to effectively match transient sound events. On the other hand, Zhang et al.~\cite{graph} use attentional graph neural networks for the same task. To address the lack of a standard benchmark in audio few-shot learning, MetaAudio~\cite{metaaudio} evaluates the most widely used few-shot learning algorithms across five publicly available datasets. Additionally, it provides predefined train-test splits, ensuring consistency in data preparation, partitioning, and backbone feature extraction.

Contrastive learning has recently gained significant attention for its effectiveness in learning robust representations. The idea is to minimize an appropriate distance metric to cluster together augmented versions (positives) of the input while distinguishing them from other samples (negatives) on the embedding space~\cite{simclr}. Building on this foundation, Khosla et al. proposed a supervised variation of the contrastive loss~\cite{supervised_contrastive} extending the original formulation by leveraging label information to pull together samples from the same class while pushing apart those from different classes. Wang et al.\cite{angular} introduced the angular loss, which improves the traditional triplet loss by enforcing a stricter angular margin between positive and negative pairs, leading to more discriminative feature representations. In few-shot image classification, several works have employed contrastive learning as an auxiliary objective \cite{spatial}, while others have integrated it directly into the training phase of few-shot models \cite{contrfew, belovedpaper}. However, in the audio domain, the integration of contrastive learning into few-shot learners remains unexplored.

To address this gap, the present work builds upon the architecture proposed in~\cite{belovedpaper}, with the aim of developing a model specifically designed for audio classification. Our approach introduces key modifications, including the integration of SpecAugment method for spectrogram augmentations~\cite{specaugment} and the replacement of the original contrastive loss with angular loss, in order to evaluate its effectiveness. For our experiments, the MetaAudio benchmark~\cite{metaaudio} is utilized to validate our approach in a 5-way, 5-shot setting across five audio datasets (ESC-50, FSDKaggle2018, VoxCeleb1, Nsynth, BirdClef2020). To facilitate the evaluation of the proposed method, the performance of the models included in the benchmark (ProtoNets \cite{Protonets}, MAML \cite{MAML}, MAML-Curvature \cite{MAMLcurva}) is also being measured. Overall, the contribution of the present work holds as follows:

\begin{itemize} 
\item[1] To the best of our knowledge, this is the first work to combine few-shot loss with supervised contrastive loss for audio classification in a few-shot setting.

\item[2] By replacing contrastive loss with angular loss, the proposed method achieves state-of-the-art results on the majority of datasets. This enables a straightforward approach such as ProtoNets to achieve competitive results compared to optimization-based algorithms, without requiring gradient updates during inference.
\end{itemize}

Finally, it is worth highlighting that this work promotes reproducibility, by thoroughly documenting all stages of our methodology and ensuring that the dataset splits, pre-processing steps, and backbone models align with those proposed in the MetaAudio benchmark. In the following GitHub repository~\href{https://github.com/magcil/audio-few-shot-learning}{https://github.com/magcil/audio-few-shot-learning} we provide instructions to reproduce the experiments, allowing further experimentation with our approach.
\section{Methods}
\label{sec:pagestyle}

\subsection{Few Shot Classification Setting}
Let $D = \left\{(x_i, y_i)\right\}_{i=1}^N$ be a collection of samples and labels. Denote by $C$ the set of all classes. Let $C = C_{\text{train}}\cup C_{\text{val}}\cup C_{\text{test}}$ be a partition of $C$. Define $D_{\text{train}} = \{(x_i,y_i)\mid y_i \in C_{\text{train}}\}$, and $D_{\text{val}}, D_{\text{test}}$, accordingly. The goal of FSL is to recognize samples from new categories by leveraging knowledge from the base training set $D_{\text{train}}$. To achieve this, FSL typically employs an episodic training strategy. In detail, $n$ classes $C_n \subset C_{train}$, and a small number of  $k +q$ samples per class are sampled from $D_{train}$ to form the support set $S = \{(x^s_i, y^s_i) \mid y^s_i \in C_n, i = 1, \hdots, n\times k \}$ and the query set $Q = \{(x^q_i, y^q_i) \mid y^q_i \in C_n, i = 1, \hdots, n\times q \}$. Together $S$ and $Q$ form an episode where $S\cap Q = \emptyset$. During training, episodes are randomly sampled from $D_{\text{train}}$; the support set provides labeled examples used as a reference for learning, while the query set consists of unlabeled samples from the same classes. The model predicts labels for the query set based on the support set, and the loss is computed using these predictions. During inference, episodes are randomly sampled from $D_{\text{test}}$ containing previously unseen classes. The model uses only a few labeled examples per class from the support set to predict the labels of query examples.

\subsection{Architecture}
Our architecture is based on~\cite{belovedpaper}, with several modifications for few-shot audio classification. The input data, instead of images, consists of single-channel mel spectrograms $x$ of shape $F\times T$, where $F$ are the frequency bins, and $T$ the time bins. The model architecture consists of four main modules: the \emph{Augmentation Module}, which generates three augmented versions of the original input; the \emph{Embedding Module} where features of each input and its augmentations are computed; the \emph{FSL Module}, where ProtoNets are used to compute the few-shot learning loss; and the \emph{Contrastive Module}, which applies two versions of supervised contrastive loss for improved representation separation.

\textbf{Augmentation Module (AM)}: To enrich the few shot batch with more information, we augment every input spectrogram $x_i^s,x_i^q$ by using time masking, frequency masking and time warping, augmentation techniques proposed by SpecAugment\cite{specaugment}. Time masking and frequency masking randomly select contiguous segments along the time and frequency axes, respectively, and mask them by setting the corresponding values to zero. Time warping applies random warping along the time axis by stretching or compressing time intervals, which helps to create more robust features. The augmentations are performed separately on each spectrogram, resulting in a list of four spectrograms: the original spectrogram and one for each applied augmentation $x_{l_i} = (x_i^{orig}, x_i^{aug_1}, x_i^{aug_2}, x_i^{aug_3}) = AM(x_i)$. 

\textbf{Embedding Module (EM)}: After the use of the AM, $x_{l_i}$ passes through the Embedding Module which is composed by a feature extraction network and a self attention module. The feature extraction network $f_{\theta}: \mathbb{R}^{F \times T} \to \mathbb{R}^D$ is a CRNN network with parameters $\theta$ that projects each element of $x_{l_i}$ to the D-dimensional feature space :
\begin{equation}
\tilde{x}_{l_i} =  [f_{\theta}(x_i^{orig}), f_{\theta}(x_i^{aug_1}),f_{\theta}(x_i^{aug_2}), f_{\theta}(x_i^{aug_3})] 
\label{xlist}
\end{equation}
The self attention module $A_{\phi}: \mathbb{R}^{4 \times D} \to \mathbb{R}^{4D}$  handles $\tilde{x_{l_i}}$ as a sequence and concatenates its output to a feature $\tilde{x}_i$ of dimension 4D:
\begin{equation}
\tilde{x}_i =  A_{\phi}([f_{\theta}(x_i^{orig}), f_{\theta}(x_i^{aug_1}),f_{\theta}(x_i^{aug_2}), f_{\theta}(x_i^{aug_3})])
\end{equation}

\textbf{Few Shot Module (FSM)}: Having both support and query inputs passed through the previous modules we get :
\begin{equation}
\begin{aligned}
\tilde{S} &= \{(\tilde{x}_i^s = EM(AM(x_i^s)), y_i^s) \mid x_i^s \in S\} \\
\tilde{Q} &= \{(\tilde{x}_i^q = EM(AM(x_i^q)), y_i^q) \mid x_i^q \in Q\}
\end{aligned}
\end{equation}

We compute class prototypes using the features in $\tilde{S}$ as shown in \eqref{prototypes}.

\begin{equation}
\tilde{p}_c = \frac{1}{k}\sum_{\tilde{x}_i, y_i \in \tilde{S}}\tilde{x}_i\cdot I(y_i = c),
\label{prototypes}
\end{equation}
 
where $I$ denotes the indicator function, returning 1 if the given condition is true, and 0 otherwise. With the class prototypes computed, we follow the approach of prototypical networks by calculating the Euclidean distance $d$ between query samples and the prototypes, and use \eqref{lfs} to compute the few shot loss $L_{fs}$.

\begin{equation}
L_{fs} = \frac{1}{q}\sum_{\tilde{x}_i, y_i \in \tilde{Q}} -\log{\frac{\exp{(-d(\tilde{x}_i, \tilde{p}_{y_i})})}{\sum_{c \in C_n}\exp{(-d(\tilde{x}_i,\tilde{p}_c))}}}
\label{lfs}
\end{equation}

\textbf{Contrastive Module (CM)}: To further improve representation separation in the embedding space, we employ two variations of the supervised contrastive loss. We begin by modifying the supervised contrastive prototype loss (CPL)~\cite{belovedpaper} by projecting both prototypes and query features through the projection head. Additionally, we employ Angular Loss, which optimizes the angular separation between embeddings rather than relying solely on Euclidean distances. In detail, the prototypes $\tilde{p}$, and the query features $\tilde{x}^q$ are passed through a small neural network $h_\beta:\mathbb{R}^{4D}\to\mathbb{R}^{D'}$ with parameters $\beta$, followed by a normalization $\hat{p} =\frac{h(\hat{p})}{\norm{h(\hat{p})}_2}, \hat{x}^q = \frac{\tilde{x}^q}{\norm{\tilde{x}^q}_2}$. The projection network $h_\beta$ allows us to experiment with various embedding dimensions $D'$ and choose the most suitable one for minimizing the final loss. We denote by $\hat{P} = \{\hat{p}_1, \dots , \hat{p}_n\}$, and $\hat{Q} = \{\hat{x}^q_1, \dots , \hat{x}^q_{qn}\}$, the sets of the projected prototypes, and queries, respectively.

\textit{\underline{Contrastive Prototype Loss (CPL)}}: The supervised contrastive loss uses the prototypes  $\hat{p}_c$, $c \in C_n$ as anchors with queries ${\hat{x}^q_c}$ of the same label forming the positive set $P_c$. We randomly sample $m$ queries from labels different from $c$ to construct the negative set $N_c$.  The loss is formulated as:

\begin{equation}
L_{cpl} = \frac{1}{nq}\sum_{c \in C_n}\sum_{(\hat{x}^q_i,y_i) \in P_c} -\log{\frac{sim_{c,i}^{(+)}}{sim_{c,i}^{(+)}+ sim_{c,i}^{(-)}} },
\label{lcp}
\end{equation}
where 
\begin{equation}
\begin{aligned}
    sim_{c,i}^{(+)} \!\!\!&= 
    \exp\frac{\inner{\hat{p_c}}{\hat{x}_{i,c}^q}}{T},\,
    sim_{c,i}^{(-)}\!\!\ = \!\!\!\!\!\!\!\!\sum_{(\hat{x}^q_t, y_t)\in N_c}\!\!\!\!\!\!\! 
    \exp\frac{\inner{\hat{p}_c}{\hat{x}_t^q}}{T}
\end{aligned}
\label{simneg}
\end{equation}

\textit{\underline{Angular Prototype Loss (APL)}}: While the CPL loss focuses on optimizing the similarity of prototypes and query pairs, the angular loss originaly proposed by \cite{angular} aims at constraining the angle at the negative point of triplet (anchor, positive, negative) triangles. Given a triplet $(x_a,x_p,x_n)$ the formulation of angular loss on a few shot batch $\mathcal{B} = \hat{P} \cup \hat{Q}$ is given by:

\begin{equation}
 \begin{aligned}
     L_{\text{apl}}(\mathcal{B})\!\! = \!\!\frac{1}{n(q+1)}\sum_{x_{\alpha}\in\mathcal{B}} \!\!\bigg\{\log\bigg[1 + \!\!\!\!\!\sum_{\substack{x_{n}\in\mathcal{B}\\ y_{n}\neq y_{a},y_{p}}}\!\!\!\!\exp\left(f_{a,p,n}\right)\bigg]\bigg\},
 \end{aligned} 
 \label{lapl}
\end{equation}

where \(f_{a,p,n}\) is defined as

\begin{equation}
f_{a,p,n} = 4\tan^{2}\alpha\inner{x_{a} + x_{p}}{x_{n}} - 2(1 + \tan^{2}\alpha)\inner{x_{a}}{x_{p}}.
\label{lapl2}
\end{equation}

The angle $\alpha\geq 0$ in~\ref{lapl2} is a predefined upper bound. The idea of angular loss is to minimize the tangent $\tan\angle n'=\frac{\norm{x_m-x_c}}{\norm{x_n-x_c}}$, where $x_c$ is the middle point of $x_n, x_p$. The point $x_m$ is one of the two points belonging on the intersection of the circle with radius $\norm{x_m-x_c}=\frac{1}{2}\norm{x_p-x_a}$, centered at $x_c$, and the hyperplane which is perpendicular to the edge $x_n-x_c$, passing through $x_c$. Minimizing~\ref{lapl2} brings $x_p, x_a$ closer on the embedding space, while pushing away the negative point $x_n$. In our case, we minimize the loss $L_{total} = L_{fs} + \lambda L_{cm}$, where $L_{cm}\in\{L_{cpl}, L_{apl}\}$, and $\lambda$ is a scaling factor. On inference time, the prototypes are derived from the set $\tilde{S}$, and the queries from $\tilde{Q}$ are classified based on their proximity to these prototypes, as in standard ProtoNets.

\section{Experiments}
\label{sec:typestyle}
\subsection{Datasets}
We follow a methodology similar to~\cite{metaaudio}, adopting the same preprocessing steps and splits for the five proposed datasets. We also reproduce the experiments of the models presented in~\cite{metaaudio} in a 5-way, 5-shot setting, under which our approach operates. ESC-50 is an environmental sound classification dataset with 2,000 clips, covering 50 different categories. FSD2018 is designed for sound event detection, featuring over 11,000 clips from 41 classes aligned with the AudioSet ontology. For musical audio, NSynth provides over 300,000 clips from 1,006 instruments, valuable for instrument recognition tasks. BirdCLEF 2020 is a bioacoustic dataset for bird species classification, offering over 80,000 recordings from 960 species. We used a pruned version of BirdCLEF 2020, removing samples longer than 180 seconds and classes with fewer than 50 samples. Finally, VoxCeleb1 serves as a speaker recognition dataset, containing utterances from various speakers in real-world conditions with background noise. We had access to a subset of VoxCeleb1, comprising 60,184 utterances from 1,246 distinct speakers, and by removing speakers with fewer than 20 recordings, we obtained 57,737 utterances from 928 speakers.

\subsection{Experimental Setup}
Audio samples from all datasets are loaded at a 16 kHz sample rate and converted to mel spectrograms. For datasets with variable-length samples (VoxCeleb1, FSD2018, BirdCLEF2020), we generate 5-second segments, as described in~\cite{metaaudio}. We apply global standardization to all spectrograms by computing the mean, and std from each training set. In all cases, the backbone is a CRNN network, consisting of a 4-block convolutional network (1-64-64-64) followed by a 1-layer non-bidirectional RNN with 64 hidden units. We train and evaluate the proposed architecture along with the ProtoNets and the optimization based models (MAML, and MAML-Curvature) presented in~\cite{metaaudio}, in a 5-way, 5-shot setting. We repeat each experiment five times and report the average accuracy and the 95\% confidence interval. In our approach, we employ a single-headed self-attention mechanism with a feedforward dimension of 256. The input is a sequence of $4\times D$, where $D=64$. The output sequence is concatenated to a $256$-dimensional embedding. We use a projection head consisting of two linear layers with hidden and output dimensions finetuned for each dataset. We conduct experiments in two different settings. In the first setting, we combine the few shot loss with the contrastive prototype loss such that $L_{total} = L_{fs} + \lambda L_{cpl}$. We denote this setting by FS+CPL. In the second setting, denoted by FS+APL we combine the few shot loss with the angular loss, i.e., $L_{total} = L_{fs} + \lambda L_{apl}$. We compute the $L_{apl}$ loss, either restricting anchors to prototypes from the support set or allowing both prototypes and queries to act as anchors. We train our models for 100 5-shot 5-way episodes per epoch over 200 epochs. We use ADAM as the optimizer, and MultiStepLR as the scheduler. We evaluate the best performing model on the validation set over 2,000 randomly sampled 5-way, 5-shot tasks from the test set. We also compare the performance of FS+CPL and FS+APL with plain ProtoNets for different number of shots (i.e., 1, 3, 5, and 7 shots). All runs were performed on an NVIDIA 4090 GPU. 

For FS+CPL we use Optuna~\cite{optuna} to determine the optimal training hyperpameters (i.e.,  $lr$, $\gamma$, $\lambda$, $T$ and $m$) based on the performance on the validation set, separately for each dataset.   
For the FS+APL setting, we use the same values for $lr$ and $\gamma$ as in the FS+CPL setup. We observe that the large values of the APL loss, compared to CPL loss, lead to increased variance among the results. We empirically find that setting $\lambda$ to a small value counteracts this effect. For this reason, we use $\lambda=0.3$ for all datasets. For the calculation of the angular loss, we use the PyTorch Metric Learning\footnote{https://kevinmusgrave.github.io/pytorch-metric-learning/} implementation. The construction of the triplets is handled by the AngularMiner where a predetermined angle threshold $\alpha$ filters-out triplets with angle less than $\alpha$, feeding harder samples to the final loss. We applied the same value of $\alpha$ for both the AngularMiner and the angular loss during each experiment, testing four different angles : 0°, 15°, 30°, and 45°. We adopt two different approaches. In the first, similar to the CPL, we use only the prototypes from the support set as anchors. In the second setting, any of the prototypes or queries can serve as anchors. We report the results of the best combination of angle, and anchor-approach for each dataset.

\subsection{Results}

\begin{table*}[t]  
\centering
\begin{tabular}{|l|c|c|c|c|c|}
\hline
\textbf{Model} & \textbf{ESC-50} & \textbf{FSD2018} & \textbf{Nsynth} & \textbf{BirdClef} & \textbf{VoxCeleb} \\
\hline
ProtoNets & 83.52 $\pm$ 0.39 & 54.19 $\pm$ 0.43 & 97.72 $\pm$ 0.17 & 71.14 $\pm$ 0.48 & 75.59 $\pm$ 0.48 \\
MAML & 87.80 $\pm$ 0.35 & 59.35 $\pm$ 0.43 & 96.73 $\pm$ 0.21 & 72.54 $\pm$ 0.48 & 73.57 $\pm$ 0.43 \\
MAML+Curv & \textbf{88.14 $\pm$ 0.30} & 57.22 $\pm$ 0.48 & \textbf{98.21 $\pm$ 0.13} & 74.30 $\pm$ 0.48 & 75.94 $\pm$ 0.43 \\
\hline
\textbf{FS+CPL} & 84.23 $\pm$ 0.35 & 58.2 $\pm$ 0.43 & 97.86 $\pm$ 0.17 & 74.95 $\pm$ 0.48 & 79.21 $\pm$ 0.48 \\
\textbf{FS+APL} & 85.61 $\pm$ 0.35 & \textbf{59.43 $\pm$ 0.43} & 97.94 $\pm$ 0.17 & \textbf{75.71 $\pm$ 0.48} & \textbf{79.78 $\pm$ 0.43} \\

APL setting & $\alpha$ = 15\textdegree ($\checkmark$) & $\alpha$ = 30\textdegree ($\times$) & $\alpha$ = 15\textdegree ($\times$) & $\alpha$ = 15\textdegree ($\times$) & $\alpha$ = 0\textdegree ($\checkmark$) \\
\hline
\end{tabular}
\caption{Performance comparison of different methods across datasets. The average accuracy and 95\% confidence interval in five runs is reported. For the FS+APL setting, we report the optimal angle threshold $\alpha$, and whether only prototypes are used as anchors $(\checkmark)$ or both prototypes and query set representations are used as anchors ($\times$).}
\label{results}
\end{table*}

\begin{figure*}[ht]
    \centering
    \includegraphics[scale=0.35]{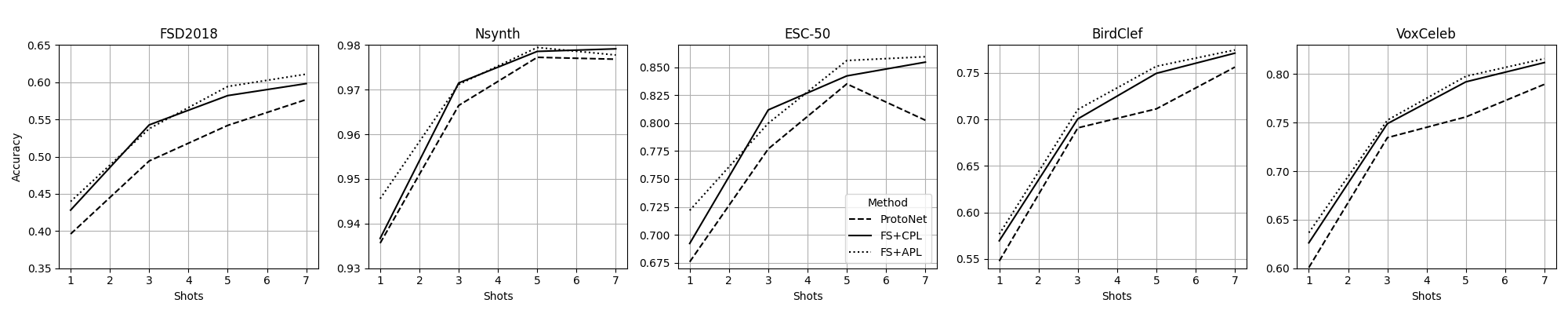}
    \caption{Comparison of ProtoNets, FS+CPL and FS+APL in different number of shots}
    \label{fig:shots}
\end{figure*}

\begin{figure}[ht]
    \centering
    \includegraphics[scale=0.6]{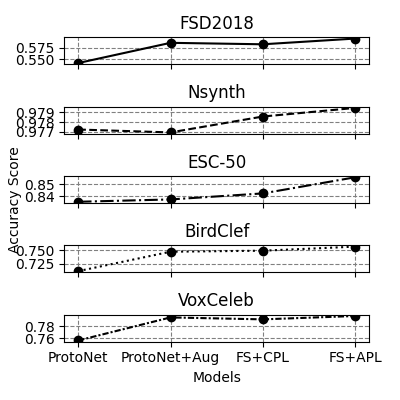}
    \caption{Module importance in overall performance per dataset.}
    \label{fig:module_comparison}
\end{figure}

Table~\ref{results} compares the performance of FS+CPL and FS+APL with the baseline architectures in~\cite{metaaudio}. We observe that both FS+CPL and FS+APL outperform ProtoNets across all datasets. In particular, FS+APL achieves significant improvements, with accuracy increases of \textbf{5.2\%} on FSD2018, \textbf{4.1\%} on VoxCeleb, \textbf{2\%} on ESC-50, \textbf{4.5\%} on BirdClef, and a slight \textbf{0.2\%} improvement on Nsynth. FS+APL also demonstrates strong performance against the optimization-based methods (MAML and MAML+Curv), surpassing the best alternative in most datasets. Specifically, on FSD2018, FS+APL matches MAML's performance with a marginal \textbf{0.08\%} accuracy increase. On BirdClef, it outperforms MAML+Curv by \textbf{1.9\%}, and on VoxCeleb it exceeds MAML+Curv by \textbf{5.6\%}. While MAML+Curv achieves slightly higher accuracy on NSynth (\textbf{0.2\%}) and ESC-50 (\textbf{2.5\%}), FS+APL remains highly competitive while requiring substantially fewer computational resources and less training time. The results highlight the effectiveness of the angular loss compared to contrastive loss and plain ProtoNets. For FS+APL, we report the best performing angle $\alpha$ separately for each dataset. However, we observed that varying the angle had a small impact on the final perfrormance. In detail, $\alpha=30^{\circ}$ yields the best results on FSD2018, $\alpha=15^{\circ}$ for Nsynth, ESC-50, and BirdClef, and $\alpha=0^{\circ}$ for VoxCeleb. Furthermore, using only prototypes as anchors improved performance on ESC-50 and VoxCeleb, while in the other datasets, the best results achieved without restricting anchors to prototypes.  Fig.~\ref{fig:shots} summarizes the performance of ProtoNets, FS+CPL and FS+APL methods across different number of shots. As it is evident, both FS+CPL and FS+APL surpass the performance of plain ProtoNets, in all datasets and all k-shot settings. As expected the performance in all datasets and models, increases with the number of shots. Overall, FS+APL performs better than FS+CPL in most k-shot scenarios. FS+CPL slightly outperforms FS+APL on the FSD2018 dataset in the 3-shot scenario by 0.51\%, and on the ESC-50 dataset in the same scenario by 1.1\%. To assess the impact of each module in our approach, we decompose it into four standalone architectures: (1) the baseline Prototypical Networks (ProtoNets); (2) ProtoNets with the augmentation module and attention layer; (3) ProtoNets with augmentation-attention and contrastive loss (FS+CPL); and (4) the same as (3) but with angular loss replacing contrastive loss (FS+APL). The 5-shot results for each dataset are presented in Fig.~\ref{fig:module_comparison}.

We observe that, except for Nsynth, the augmentation-attention module improves the accuracy of the plain Prototypical Networks across all datasets. Specifically, this module increases accuracy by 4.32\% in FSD2018, 3.63\% in BirdClef, 3.94\% in VoxCeleb, and 0.20\% in ESC-50. In Nsynth, however, the accuracy exhibits a very slight decrease of 0.03\%.
The inclusion of Contrastive loss (FS+CPL) further enhances accuracy in most datasets. Compared to the augmentation-attention module alone, it adds 0.52\% in ESC-50 and 0.16\% in Nsynth, while showing a minor decrease of 0.31\% in FSD2018 and 0.33\% in VoxCeleb. In BirdClef, the improvement is 0.19\%.
By replacing Contrastive loss with Angular loss (FS+APL), we achieve further performance improvements over the augmentation-attention module. Specifically, FS+APL increases accuracy by 0.91\% in FSD2018, 0.25\% in Nsynth, 1.90\% in ESC-50, 0.95\% in BirdClef, and 0.24\% in VoxCeleb compared to the augmentation-attention module.

\section{Conclusions}

We presented a novel approach for few-shot audio classification that enhances ProtoNets utilizing spectrogram augmentation and contrastive learning. Overall, our work is the first to integrate supervised contrastive learning, specifically angular loss, into prototypical few-shot training for audio classification. Extensive evaluation on the MetaAudio benchmark demonstrates state-of-the-art performance in 5-way 5-shot classification, showing significant improvements over standard ProtoNets (up to 5.2\% on challenging datasets) while matching the accuracy of more computationally intensive optimization-based approaches. Future research directions include investigating alternative contrastive loss formulations and developing more sophisticated training techniques to further boost few-shot learning performance.

\bibliographystyle{IEEEbib}
\bibliography{strings,refs}

\end{document}